\documentclass{article}
\usepackage[utf8]{inputenc}
\usepackage[T1]{fontenc} 
\usepackage{arxiv}
\usepackage{amsmath}
\usepackage{braket}
\usepackage{amsfonts}
\usepackage{amssymb}
\usepackage{graphicx}
\usepackage{amsthm}
\usepackage{listings}
\usepackage{float}
\usepackage{caption}
\usepackage{subcaption}
\usepackage{color}
\usepackage[colorlinks=true,allcolors=blue]{hyperref}
\usepackage{chngcntr}
\usepackage{cite}
\usepackage{authblk}
\usepackage{tikz}
\usepackage{booktabs} 
\DeclareFontFamily{OMS}{oasy}{\skewchar\font48 }
\DeclareFontShape{OMS}{oasy}{m}{n}{%
	<-5.5> oasy5     <5.5-6.5> oasy6
	<6.5-7.5> oasy7     <7.5-8.5> oasy8
	<8.5-9.5> oasy9     <9.5->  oasy10
}{}
\DeclareFontShape{OMS}{oasy}{b}{n}{%
	<-6> oabsy5
	<6-8> oabsy7
	<8->  oabsy10
}{}
\DeclareSymbolFont{oasy}{OMS}{oasy}{m}{n}
\SetSymbolFont{oasy}{bold}{OMS}{oasy}{b}{n}

\DeclareMathSymbol{\smallleftarrow}     {\mathrel}{oasy}{"20}
\DeclareMathSymbol{\smallrightarrow}    {\mathrel}{oasy}{"21}
\DeclareMathSymbol{\smallleftrightarrow}{\mathrel}{oasy}{"24}

\newcommand{\tensor}[1]{\overset{\scriptscriptstyle\smallleftrightarrow}{#1}}

\definecolor{greennew}{rgb}{0.07, 0.53, 0.37}

\counterwithout{figure}{section}

\title{Coherent states for dispersive pseudo-Landau-levels in strained honeycomb lattices}

\author{Erik D\'iaz-Bautista\textsuperscript{1} and Maurice Oliva-Leyva\textsuperscript{2}\\
\textsuperscript{1}Departamento de Formaci\'on B\'asica Disciplinaria, Unidad Profesional Interdisciplinaria de Ingenier\'ia Campus Hidalgo del Instituto Polit\'ecnico Nacional, Pachuca: Ciudad del Conocimiento y la Cultura, Carretera Pachuca-Actopan km 1+500, 42162 San Agust\'in Tlaxiaca, Hidalgo, Mexico.\\
\textsuperscript{2}Departamento de F\'isica, Facultad de Ciencias, Universidad Nacional Aut\'onoma de M\'exico, 04510 Mexico City, Mexico.\\
e-mail: ediazba@ipn.mx, moliva@ciencias.unam.mx}
\date{ }

\begin{document}

\maketitle
\begin{abstract}
    Dirac fermions in graphene may experiment dispersive pseudo-Landau levels due to a homogeneous pseudomagnetic field and a position-dependent Fermi velocity induced by strain. In this paper, we study the (semi-classical) dynamics of these particles under such a physical context from an  approach of coherent states. For this purpose we use a Landau-like gauge to built Perelomov coherent states by the action of a non-unitary displacement operator $D(\alpha)$ on the fundamental state of the system. We analyze the time evolution of the probability density and the generalized uncertainty principle as well as the Wigner function for the coherent states. Our results show how $x$-momentum dependency affects the motion periodicity and the Wigner function shape in phase space.
\end{abstract}

\section{Introduction}\label{sec1}
The harmonic oscillator is one of the most important physical systems since any smooth potential can be approximated as a harmonic oscillator potential close to an equilibrium point. In quantum mechanics, the harmonic oscillator model constitutes a useful theoretical tool that allows describing successfully many quantum phenomena, e.g., electromagnetic field quantization and lattice and molecular vibrations. This system is characterized by its equidistant discrete spectrum and its nonzero minimum energy level. Also, as energy increases, the excited state probability density shows peaks at the classical returning points. This fact motivated Schr\"{o}dinger~\cite{s26} to define a type of minimal uncertainty quantum states whose wave packet described the classical motion of a particle in a square potential in agreement to the correspondence principle. Then, Glauber~\cite{g63} rediscovered such states in order to describe the electromagnetic field, naming them as coherent states (CSs). Coherent states can be obtained for the harmonic oscillator through three different definitions, which are mathematically equivalent to each other and that have been generalized to obtain coherent states with complex dynamical properties~\cite{klauder85,Manko1997,Gazeau1999,gazeau10,Recamier2008}: 1) as states with minimal uncertainty~\cite{s26,nieto78,nieto79}, 2) as the right-hand eigenstates of the annihilation operator, called in literature as Barut–Girardello coherent states~\cite{g63,glauber63,Barut1971}, and 3) as states obtained by the application of the displacement operator $D(\alpha)$ on the ground state of the harmonic oscillator, known as Gilmore-Perelomov coherent states~\cite{klauder63,perelomov72,Gilmore1974}.

With regard to condensed matter systems, the coherent state formalism has been considered, for instance, in the theory of superconductivity~\cite{anderson58,HUANGHONG1991,SHCHUROVA2004}. Following this trend, the study of CSs has been recently extended to low dimensional materials. Specifically, with the arrival of graphene and the new family of Dirac materials, whose electrons behave in many aspects as relativistic particles that are governed by the Dirac equation rather than by the Schr\"{o}dinger equation, and since CSs have been widely studied in the context of a non-relativistic charged particle under an uniform magnetic field~\cite{fock28,landau30,feldman70,Dodonov2018}, because this problem (known as Landau problem) can be mapped to the harmonic oscillator problem, Barut–Girardello coherent states have been built for describing the interaction between electrons in monolayer graphene and external electric and magnetic fields~\cite{df17,Diaz4,celeita2020}, in order to describe the dynamics of the so-called Dirac fermions from a semi-classical approach. Besides, CSs have also been obtained for bilayer graphene interacting with a homogeneous magnetic field~\cite{mmfc20}.

On the other hand, among the most interesting features of graphene, one can cite the peculiar interplay between its electronic and its mechanical properties. Given its unusual long interval of elastic response (up to $20\%$), strain engineering has been exploited to tailor its electronic and optical properties and, thus, its optoelectronic functionalities~\cite{Naumis2017,Peng2020}. For instance, under uniform strains, the graphene electronic band structure around the Dirac points becomes elliptical cones, which is traduced in an anisotropy of the Fermi velocity \cite{Oliva2017}, and, as a consequence, the optical conductivity of uniformly deformed graphene results anisotropic \cite{Pellegrino2010}. This fact produces a strain-induced modulation of the optical transmittance \cite{Ni2014} and of the Faraday (Kerr) effect in graphene \cite{Oliva2017b}. 
Moreover, it is precisely by taking advantage of the strain-induced anisotropy that the concept of strain engineering has been also extended to the context of CSs in recent works \cite{dcr19,diazbetancur20,diaz2019coherent,dbe20,Anbaraki2021}, but all limited to the case of uniform strains.

However, unlike uniform strains, nonuniform strains constitute a more efficient tool to investigate new and striking behaviors of graphene such as superconducting states \cite{Kauppila2016,Mao2020} or Hall effect even in absence of external magnetic field \cite{Sela2020,Wagner2020}. This is due mainly to the emergence of a pseudomagnetic field caused by nonuniform strains. Nowadays, signatures of such pseudomagnetic field  in the electronic transport properties of graphene are actively explored, particularly those related with valleytronic \cite{Settnes2016,Stegmann2018,He2020,Lantagne2020}. For example, Lantagne et al.~\cite{Lantagne2020} recently drew attention to the possibility of achieving spatially separated valley currents in graphene nanoribbons subjected to uniaxial nonuniform strain. This finding is based on the following remarkable fact: In the presence of certain uniform pseudomagnetic fields in strained graphene, the resulting pseudo-Landau-levels are not flat but disperse, i.e. dependent on the wave vector. Such dispersive behavior of the pseudo-Landau-levels in strained graphene has been explained in terms of a position-dependent Fermi velocity \cite{Lantagne2020,maurice2020}, which is also another known effect induced by nonuniform strains \cite{FJ2012,Oliva2015a}.

It is just this scenario of dispersive pseudo-Landau-levels in strained honeycomb lattices that motives the present study of CSs. In this regard, here we consider a different expression for the annihilation operator $\Theta^{-}$, in comparison with previous works, in order to apply an alternative definition to obtain the CSs, such as the displacement-operator method, taking advantage of the lack of uniqueness of the form of the operator $\Theta^{-}$. So, our main aim consists of improving the construction of CSs in 2D Dirac materials for describing the dynamics of charge carriers in condensed matter systems from a semi-classical approach, through the time evolution of the coherent state wave packet, the definition of quadratures and phase-space representation.

Thus, this paper is organized as follows. In Section~\ref{sec2} we briefly discuss the effective Dirac Hamiltonian with a $x$-position-dependent Fermi velocity. We also show the corresponding spectrum and eigenvectors. In Section~\ref{sec3} we define the matrix ladder operators associated with the physical system, which allow us to obtain a non-unitary displacement operator. We built the Perelomov coherent states by applying such a displacement operator on the fundamental eigenvector. We also study the time evolution of the coherent states through the probability density, the corresponding auto-correlation function, as well as the generalized uncertainty principle. In Section~\ref{sec4} we obtain the phase-space representation of the coherent states through the corresponding Wigner function. This function allows us to analyze the dynamics in phase space. In Section~\ref{conclusions} we comment our conclusions and final remarks.

\section{The model}\label{sec2}

\begin{figure}[ht]
	\centering
	\includegraphics[width=14cm]{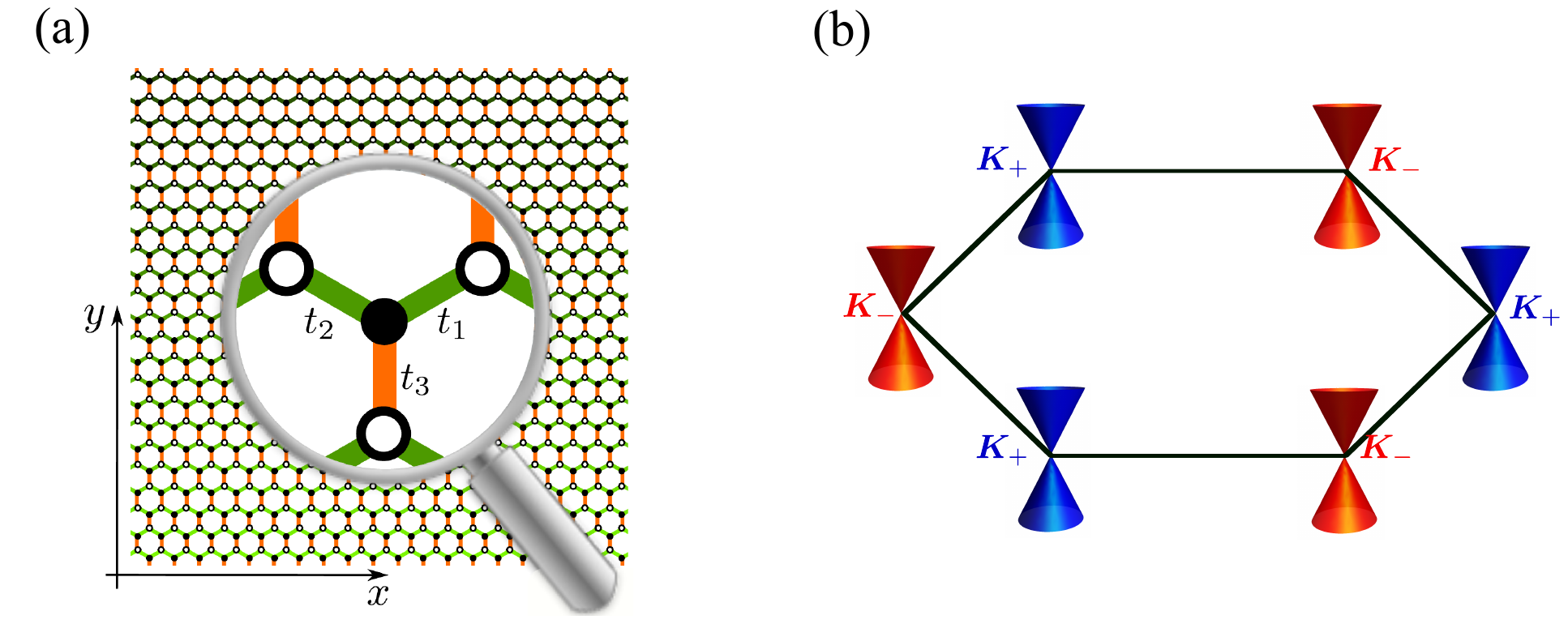}\caption{\label{fig:Red} (a) Illustration of an anisotropic honeycomb lattice where the nearest-neighbor hoppings $t_1$ and $t_2$ (in green color) are dependent on the $y$-coordinate. This fact is displayed with the change of the green hue along of $y$-direction. Otherwise, the hopping $t_3$ (in orange color) is constant. (b) Representation of the two (inequivalent) Dirac valleys at the vertices of the Brillouin zone for an isotropic honeycomb lattice.}
\end{figure}

Here we consider the effective model for the low-energy quasiparticles in an anisotropic honeycomb lattice (e.g. strained graphene), where the hopping parameters between nearest sites ($t_1$ , $t_2$ and $t_3$) vary along the $y$-coordinate as
\begin{align}\label{hps}
    t_1(y)=t_2(y)=t(1+3c_{0}y/4), \quad t_{3}=t,
\end{align}
$t$ being the hopping parameter of the unperturbed lattice (see fig. \ref{fig:Red}(a)). Note that, in the framework of tight-binding model, the hopping parameters are fundamental ingredients that capture the possibility of the electrons to hop between neighboring lattice sites \cite{Simon}. For instance, in the extreme case of $t\rightarrow 0$, it is impossible for an electron to hop into neighboring sites. Otherwise, for graphene $t\approx2.7\,\text{eV}$ which is the value used in our calculations. 

In this scenario the following effective Dirac Hamiltonian arises~\cite{maurice2020}
\begin{equation}\label{H}
    H=v_{\rm F}\left([(1+c_{0}y)p_{x}+eB_{\rm ps}y]\sigma_{x}+p_{y}\sigma_{y}\right),
\end{equation} 
where $\sigma_{x,y}$ are the Pauli matrices,  $v_{\rm F}=3at/2\hbar$ is the Fermi velocity of the unperturbed lattice and $B_{\rm ps}=\tau\hbar\,c_{0}/(2ea)$ is the strength of the pseudomagnetic field induced by the variation of the hopping parameters (\ref{hps}),  with $\tau$ being the valley index which takes the values $+1$ or $-1$ and specifies whether Eq.~(\ref{H}) is obtained by an expansion around the points of the Brillouin zone, $\mathbf{K}_{+}$ or $\mathbf{K}_{-}$, respectively (see fig. \ref{fig:Red}(b)).  

It is worth noting that in addition to the uniform pseudomagnetic field, the hopping variation (\ref{hps}) also induces a position-dependent Fermi velocity whose anisotropic character can be captured by the tensor~\cite{maurice2020}
\begin{equation}\label{tensorvelocity}
    \tensor{v}=v_{\rm F}\left(\begin{array}{cc}
        1+c_{0}y & 0 \\
        0 & 1
    \end{array}\right).
\end{equation}
which expresses that the Fermi velocity in the $x$-direction depends linearly on the $y$-coordinate. 

According to the translational symmetry of (\ref{H}) along the $x$-direction, the eigenfunctions can be represented $\Psi(\mathbf{r})=\exp\left(ik_{x}x\right)\left(\psi_{1}(y) \quad \psi_{2}(y)\right)^{\rm T}$ (where ${\rm T}$ denotes transpose) and, as a consequence, the eigenvalue equation $H\Psi(\mathbf{r})=E\Psi(\mathbf{r})$ can be rewritten as
\begin{equation}\label{eigeneq}
    \hbar\,v_{\rm F}\left(\left[k_{x}+\frac{e\mathcal{B}}{\hbar}y\right]\sigma_{x}-i\partial_{y}\sigma_{y}\right)\left(\begin{array}{c}
      \psi_{1}(y) \\
      \psi_{2}(y)   
    \end{array}\right)=E\left(\begin{array}{c}
      \psi_{1}(y) \\
      \psi_{2}(y)   
    \end{array}\right),
\end{equation}
where $E$ is the energy and $\mathcal{B}=B_{\rm ps}(1 + \tau\,2k_{x}a)$ whose meaning is clarified below.

Now, by defining the quantities
\begin{equation}
    \xi=\frac{1}{l_{\rm B}}(y+l_{\rm B}^{2}k_{x}), \quad l_{\rm B}^{2}=\frac{\hbar}{e\mathcal{B}},
\end{equation}
the expression in~(\ref{eigeneq}) yields to
\begin{equation}\label{coupledeq}
    \left(\begin{array}{cc}
      0  & \theta^{+} \\
      \theta^{-} & 0
    \end{array}\right)\left(\begin{array}{c}
      \psi_{1}(y) \\
      \psi_{2}(y)   
    \end{array}\right)=\frac{E\,l_{\rm B}}{\sqrt{2}\hbar\,v_{\rm F}}\left(\begin{array}{c}
      \psi_{1}(y) \\
      \psi_{2}(y)   
    \end{array}\right),
\end{equation}
where
\begin{equation}
    \theta^{\pm}=\frac{1}{\sqrt{2}}\left(\xi\mp\frac{d}{d\xi}\right)
\end{equation}
are scalar differential operators that satisfy the commutation relation $[\theta^{-},\theta^{+}]=1$.

By simple inspection, one can see that Eq.~(\ref{coupledeq}) is analogous to the Landau level (LL) problem for massless Dirac fermions in the presence of an effective homogeneous magnetic field of strength $\mathcal{B}$. Thus, it is straightforward to verify that the spectrum is given by
\begin{equation}\label{LLs}
    E_{n,k_{x}}=\lambda\,v_{\rm F}\sqrt{2\hbar\,e\vert\mathcal{B}\vert n}=\lambda\hbar\,v_{\rm F}\sqrt{\vert c_{0}/a\vert n}\sqrt{1+\tau\,2k_{x}a},
\end{equation}
with $\lambda=\pm$ being the band index. Unlike the standard Landau levels, the resulting levels (\ref{LLs}) are \textit{dispersive}, as coined by Lantagne et al. \cite{Lantagne2020}, because they depend on $k_x$. For small $k_x$, the dependence is linear with an opposite slope sign for each valley.

Meanwhile, the corresponding normalized eigenvectors can be expressed as
\begin{equation}\label{eigenvectors}
    \Psi_{n}(\mathbf{r})=\frac{\exp\left(ik_{x}x\right)}{\sqrt{2^{(1-\delta_{0n})}}}\left(\begin{array}{c}
        \lambda\,\tau\,\psi_{n}(y) \\
        (1-\delta_{0n})\psi_{n-1}(y)  
    \end{array}\right),
\end{equation}
where $\delta_{mn}$ denotes the Kronecker delta,
\begin{equation}\label{wavefunctions}
    \psi_{n}(y)=\sqrt{\frac{1}{2^{n}n!\,l_{\rm B}\sqrt{\pi}}}H_{n}\left(\frac{1}{l_{\rm B}}\left(y+l_{\rm B}^{2}k_{x}\right)\right)\exp\left(-\frac{1}{2l_{\rm B}^{2}}\left(y+l_{\rm B}^{2}k_{x}\right)^{2}\right),
\end{equation}
and $H_{k}(z)$ indicates the Hermite polynomial of degree $k$. It is worth to mention that the quantity $\omega=v_{\rm F}\sqrt{\vert c_{0}/a\vert}$ is the cyclotron-like frequency for the Dirac fermions.

As mentioned before, Eqs.~(\ref{eigeneq})-(\ref{wavefunctions}) correspond to the physical system in which an external homogeneous magnetic field $\mathbf{B}_{\rm ext}=-\mathcal{B}_{\rm ext}\mathbf{\hat{k}}$ is applied to the sample, and that can be described by a Landau-like gauge with translational invariance along the $y$-axis, namely, $\mathbf{A}(y)=\mathcal{B}_{\rm ext}y\mathbf{\hat{i}}$. So, we shall also consider such a case by replacing $\mathcal{B}\rightarrow\mathcal{B}_{\rm ext}$. Also, without loss of generality, in the forthcoming sections we will focus in the conduction band ($\lambda=+$).

\section{Perelomov coherent states}\label{sec3}
Let us consider the following matrix operators
\begin{equation}
    \Theta^{-}=\frac{1}{\sqrt{2}}\left(\begin{array}{cc}
    \theta^{-} & \tau\,\sqrt{N+1} \\
   \tau\,\frac{1}{\sqrt{N+1}}(\theta^{-})^{2} & \frac{\sqrt{N+2}}{\sqrt{N+1}}\theta^{-}
    \end{array}\right), \quad \Theta^{+}=\frac{1}{\sqrt{2}}\left(\begin{array}{cc}
    \theta^{+} & \tau\,(\theta^{+})^{2}\frac{1}{\sqrt{N+1}} \\
    \tau\,\sqrt{N+1} & \theta^{+}\frac{\sqrt{N+2}}{\sqrt{N+1}}
    \end{array}\right),
\end{equation}
such that $\Theta^{+}=(\Theta^{-})^{\dagger}$, where $N\equiv\theta^{+}\theta^{-}$, and whose actions on the eigenvectors in Eq.~(\ref{eigenvectors}) read as
\begin{equation}
    \Theta^{-}\Psi_{n}(\mathbf{r})=\sqrt{2^{(1-\delta_{1n})}}\sqrt{n}\Psi_{n-1}(\mathbf{r}), \quad \Theta^{+}\Psi_{n}(\mathbf{r})=\sqrt{2^{(1-\delta_{0n})}}\sqrt{n+1}\Psi_{n+1}(\mathbf{r}), \quad n=0,1,2,\dots.
\end{equation}
Also, these matrix operators satisfy the commutation relation
\begin{equation}
    [\Theta^{-},\Theta^{+}]\Psi_{n}=c(n)\Psi_{n}, \quad c(n)=\begin{cases}
    1, & n=0, \\
    3, & n=1, \\
    2, & n>1.
    \end{cases}
\end{equation}

Thus, we are able to obtain excited states from the fundamental one $\Psi_{0}$ as follows
\begin{equation}
    \Psi_{n}=\frac{\sqrt{2^{(1-n-\delta_{0n})}}}{\sqrt{n!}}(\Theta^{+})^{n}\Psi_{0}, \quad n=0,1,2,\dots,
\end{equation}
i.e., the matrix operator $\Theta^{+}$ works, up to a constant, as a creation operator for the Hilbert space $\mathcal{H}$.

\subsection{Obtaining of coherent states}
In order to construct the coherent states, we define a displacement operator $D(\alpha)$ as follows:
\begin{equation}
    D(\alpha)=\exp\left(\alpha\Theta^{+}\right)\exp\left(\alpha^{\ast}\Theta^{-}\right).
\end{equation}

Thus, the Perelomov coherent states (PCSs) are built by acting the displacement operator on the ground state as follows:
\begin{equation}
\Psi_{\alpha}=D(\alpha)\Psi_{0}=\exp\left(\alpha\Theta^{+}\right)\exp\left(\alpha^{\ast}\Theta^{-}\right)\Psi_{0}=\sum_{n=0}^{\infty}\frac{\alpha^{n}}{n!}(\Theta^{+})^{n}\Psi_{0}=\Psi_{0}+\frac{1}{\sqrt{2}}\sum_{n=1}^{\infty}\frac{(\sqrt{2}\alpha)^{n}}{\sqrt{n!}}\Psi_{n}.
\end{equation}
These states must be normalized since the operator $D(\alpha)$ is not unitary.

\begin{figure}[ht]
	\centering
	\includegraphics[width=0.8\textwidth]{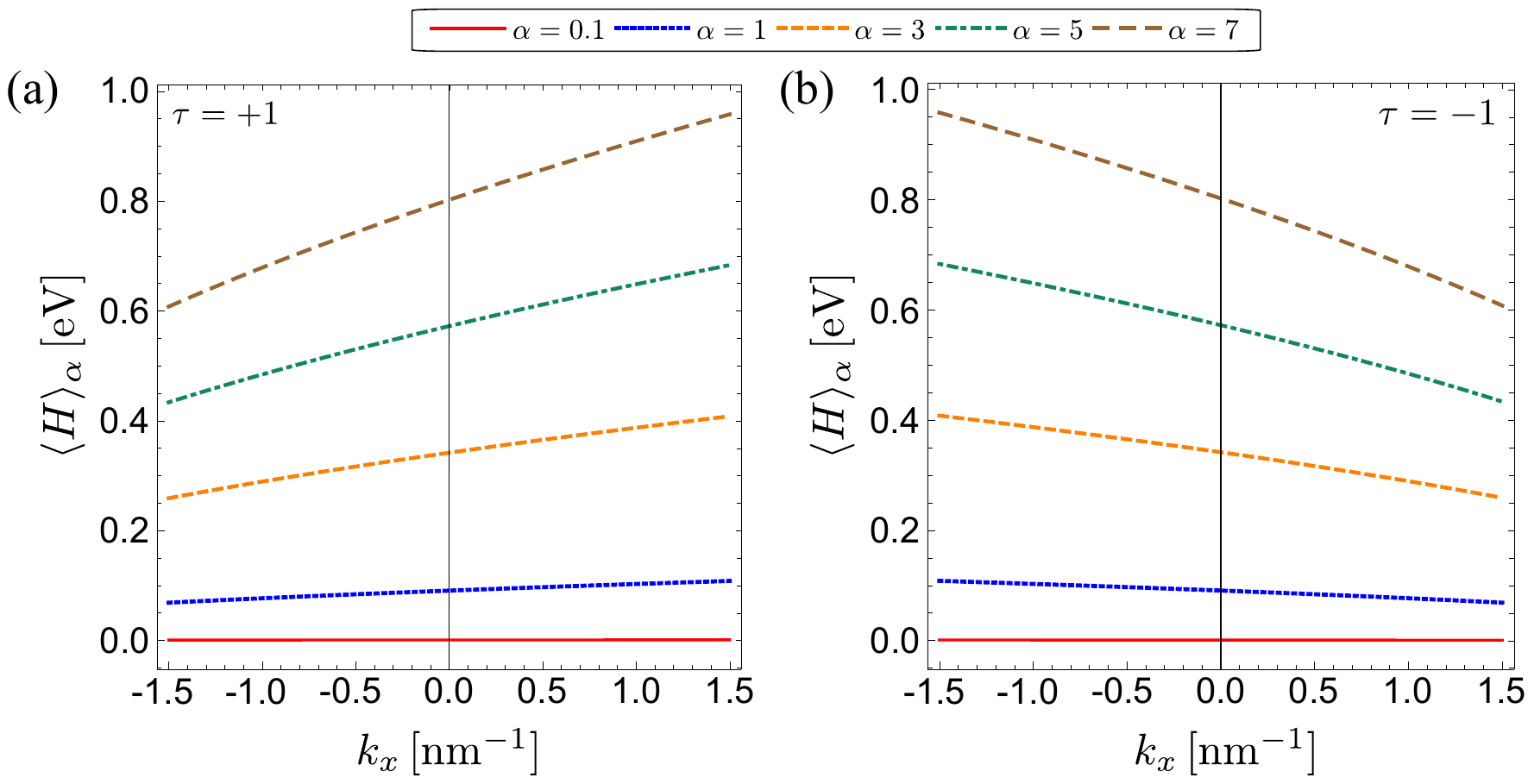}
	\caption{\label{fig:energy} Mean energy value $\langle H\rangle_{\alpha}$ with $B_{\rm ps}=5$ T and different values of $\vert\alpha\vert$ as function of $k_{x}$ for the valleys $\mathbf{K}_{+}$ (a) and $\mathbf{K}_{-}$ (b).}
\end{figure}

The normalized PCSs are
\begin{equation}\label{PCSs}
\Psi_{\alpha}=\sqrt{\frac{2}{\left(\exp\left(\vert z\vert^{2}\right)+1\right)}}\left[\Psi_{0}+\frac{1}{\sqrt{2}}\sum_{n=1}^{\infty}\frac{z^{n}}{\sqrt{n!}}\Psi_{n}\right]=\frac{\exp\left(ik_{x}x\right)}{\sqrt{2\left(\exp\left(\vert z\vert^{2}\right)+1\right)}}\left(\begin{array}{c}
      \tau\,(\psi_{0}+\psi_{\alpha})  \\
       \psi'_{\alpha}
    \end{array}\right),
\end{equation}
where $z=\sqrt{2}\alpha$ and
\begin{subequations}\label{components}
\begin{align}
    \psi_{\alpha}&=\left(\frac{1}{\sqrt{\pi}l_{\rm B}}\right)^{1/2}\exp\left(-\frac{\xi^{2}}{2}-\frac{z^{2}}{2}+\sqrt{2}z\xi\right), \label{componentsa} \\ 
    \psi'_{\alpha}&=\left(\frac{1}{\sqrt{\pi}l_{\rm B}}\right)^{1/2}\exp\left(-\frac{\xi^{2}}{2}\right)\sum_{n=1}^{\infty}\frac{(z/\sqrt{2})^{n}}{n!}\sqrt{2n}H_{n-1}(\xi). \label{componentsb}
\end{align}
\end{subequations}
Above, Eq.~(\ref{componentsa}) corresponds to the wave function of an unnormalized standard coherent state. Also, the corresponding probability density read as
\begin{equation}\label{densityCS}
\vert\Psi_{\alpha}\vert^{2}=\frac{1}{2\left(\exp\left(\vert z\vert^{2}\right)+1\right)}\left[\vert\psi_{0}+\psi_{\alpha}\vert^{2}+\left\vert\psi'_{\alpha}\right\vert^{2}\right].
\end{equation}

\subsection{Mean energy value}
On the other hand, the mean energy value $\langle H\rangle_{\alpha}$ is given by (see Fig.~\ref{fig:energy})
\begin{equation}\label{meanenergy}
    \langle H\rangle_{\alpha}=\frac{v_{\rm F}\sqrt{\hbar\,eB_{\rm ps}(\tau\,2k_{x}a+1)}}{\left(\exp\left(\vert z\vert^{2}\right)+1\right)}\sum_{n=1}^{\infty}\frac{\vert z\vert^{2n}}{n!}\sqrt{2n}.
\end{equation}

Recalling the case in which an external homogeneous magnetic field $\mathbf{B}_{\rm ext}$ is applied to the graphene layer on $xy$-plane, and since in this one the energy spectrum is independent of the momentum $k_{x}$ and the valley index $\tau$, one can easily compare the mean energy value in both cases. As we can see in Fig.~\ref{fig:energy}, depending on which valley is considered, the slope of the function $\langle H\rangle_{\alpha}$ changes.

\begin{figure}[ht]
	\centering
	\includegraphics[width=0.98\textwidth]{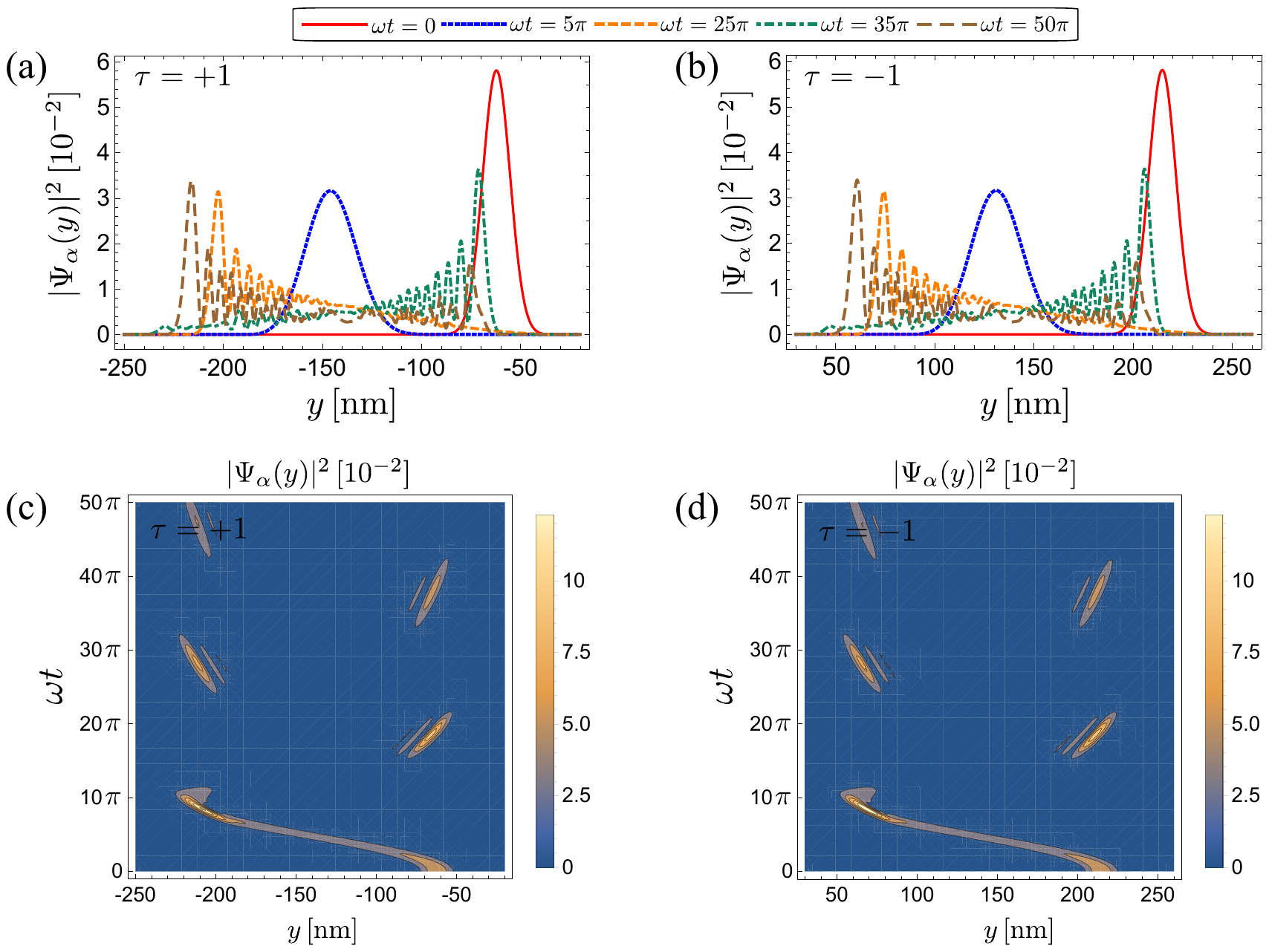}
	\caption{\label{fig:density}Time-dependent probability density $\vert\Psi_{\alpha}(y,t)\vert^{2}$ with $\alpha=4$ and $B_{\rm ps}=5$ T as function of $\omega\,t$. (a, c) For the valley $\mathbf{K}_{+}$ with $k_{x}=1.5$ nm$^{-1}$ and (b, d) for the valley $\mathbf{K}_{-}$ with $k_{x}=-1.5$.}
\end{figure}

\subsection{Time evolution}
Now, we investigate the time evolution of the PCSs by applying the time evolution unitary operator $U(t,t_{0})=\exp(-i H(t-t_{0})/\hbar)$ on the states $\Psi_{\alpha}$ in Eq.~(\ref{PCSs}):
\begin{equation}
    \Psi_{\alpha}(t)=U(t,t_{0})\Psi_{\alpha}, \quad t>t_{0},
\end{equation}
such that $U(t_{0},t_{0})=\mathbb{I}$, being the $\mathbb{I}$ identity operator. Thus, setting $t_{0}=0$, we have
\begin{equation}\label{PCSstime}
    \Psi_{\alpha}(t)=\sqrt{\frac{2}{\left(\exp\left(\vert z\vert^{2}\right)+1\right)}}\left[\Psi_{0}+\frac{1}{\sqrt{2}}\sum_{n=1}^{\infty}\frac{ z^{n}}{\sqrt{n!}}\exp\left(-\frac{i E_{n,k_{x}}t}{\hbar}\right)\Psi_{n}\right],
\end{equation}
while the corresponding probability density is
\begin{equation}\label{densityCStime}
\vert\Psi_{\alpha}(t)\vert^{2}=\frac{1}{2\left(\exp\left(\vert z\vert^{2}\right)+1\right)}\left[\left\vert\psi_{0}+\sum_{n=1}^{\infty}\frac{ z^{n}}{\sqrt{n!}}\exp\left(-\frac{i  E_{n,k_{x}}t}{\hbar}\right)\psi_{n}\right\vert^{2}
+\left\vert\sum_{n=1}^{\infty}\frac{ z^{n}}{\sqrt{n!}}\exp\left(-\frac{i  E_{n,k_{x}}t}{\hbar}\right)\psi_{n-1}\right\vert^{2}\right].
\end{equation}

As we can see in Fig.~\ref{fig:density}, the time evolution of the coherent state wave function is periodic, being such a period $T$ dependent of the valley index. However, due to the non-equidistant spectrum of the system, the shape of the probability density is not stable in time: as $t$ increases, the function $\vert\Psi_{\alpha}(t)\vert^{2}$ varies, showing many oscillations while its amplitude decreases. It is worth to remark that the probability density shows maximum values in the classical returning points at $T/2$, which agrees with the fact that the particle velocity is momentarily zero at returning points. Likewise, $\alpha$ is a complex parameter whose phase establishes the initial conditions of the harmonic motion that the charge carriers perform, while $\vert\alpha\vert$ gives the amplitude of the oscillations~\cite{Cohen}.

\begin{figure}[ht]
	\centering
	\includegraphics[width=0.98\textwidth]{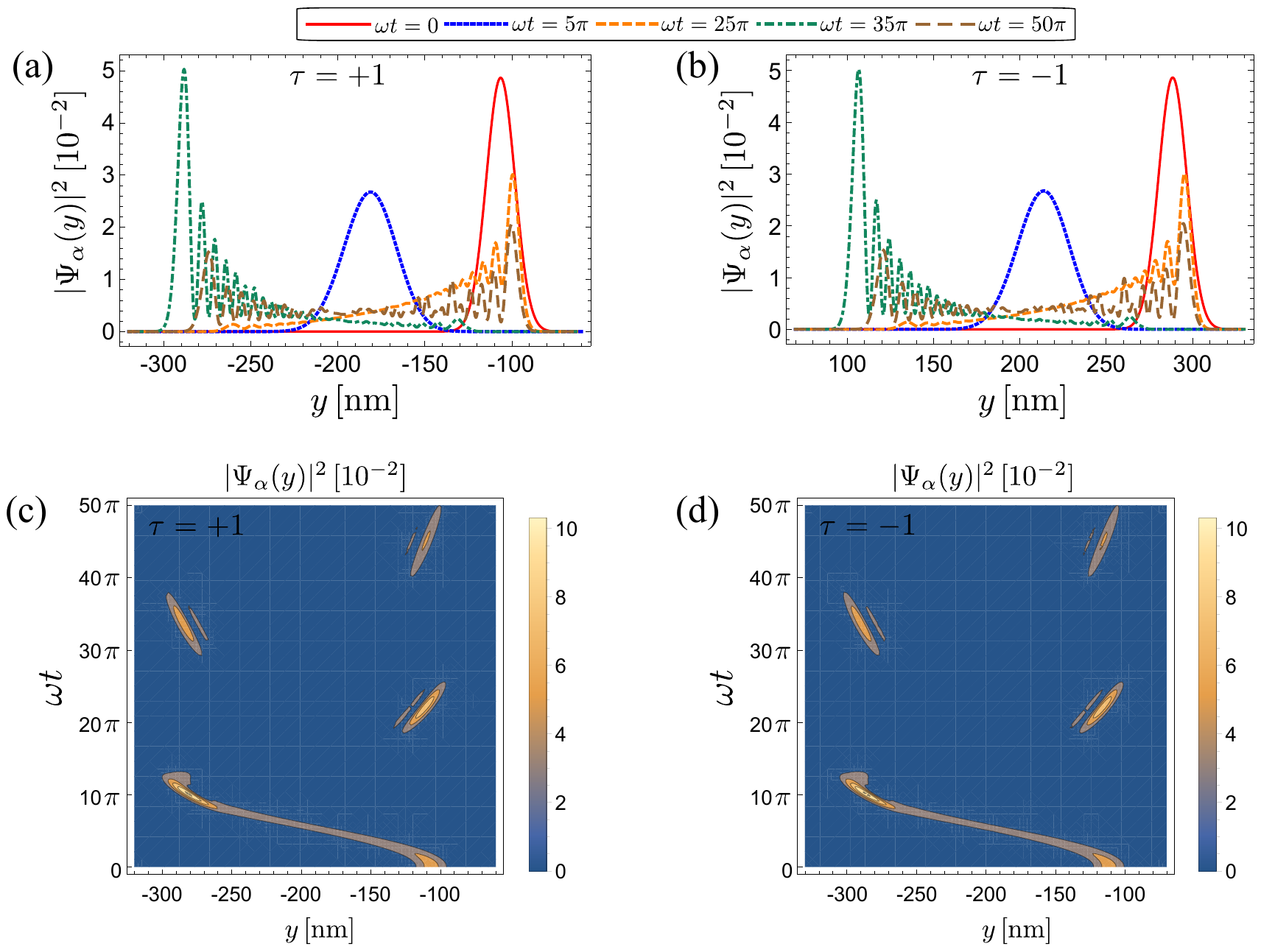}
	\caption{\label{fig:comparison1}For comparison: time-dependent probability density $\vert\Psi_{\alpha}(y,t)\vert^{2}$, with $\alpha=4$, as function of $\omega\,t$ and for an external uniform magnetic of strength $\mathcal{B}_{\rm ext}=5$ T applied to a pristine graphene sample. (a, c) For the valley $\mathbf{K}_{+}$ with $k_{x}=1.5$ nm$^{-1}$ and (b, d) for the valley $\mathbf{K}_{-}$ with $k_{x}=-1.5$.}
\end{figure}

In contrast with the case in which an external homogeneous magnetic field $\mathbf{B}_{\rm ext}$ is applied to the graphene layer, we can note that the corresponding time evolution of the probability density, shown in Fig.~\ref{fig:comparison1}, has a different period. Being more precise, for the same values of the parameter $\alpha$, magnetic field strength and momentum $k_{x}$, the $x$-momentum-independent case possess a major motion period, which allows observing revivals in the probability density in time. Also, the point along the $y$-axis in which the particle locates at $t=0$ is modified due to the $x$-momentum dependency of the eigenfunctions $\psi_{n}$ in (\ref{wavefunctions}).

\begin{figure}[ht]
	\centering
	\includegraphics[width=0.9\textwidth]{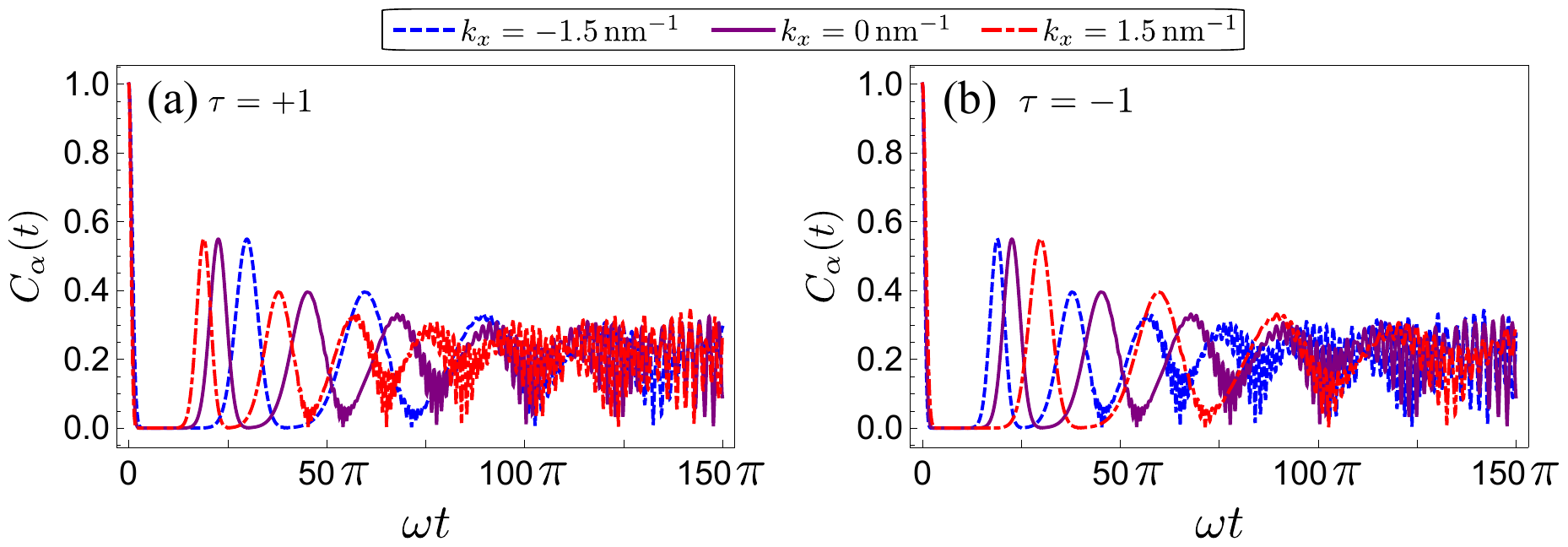}
	\caption{\label{fig:autocorrelation}Auto-correlation function $C_{\alpha}(t)$ with $\alpha=4$, as function of $\omega\,t$, for $B_{\rm ps}=5\,\text{T}$ and three different values of $k_{x}$: (a) for the valley $\mathbf{K}_{+}$ and (b) for the valley $\mathbf{K}_{-}$.}
\end{figure}

\begin{figure}[ht]
	\centering
	\includegraphics[width=0.9\textwidth]{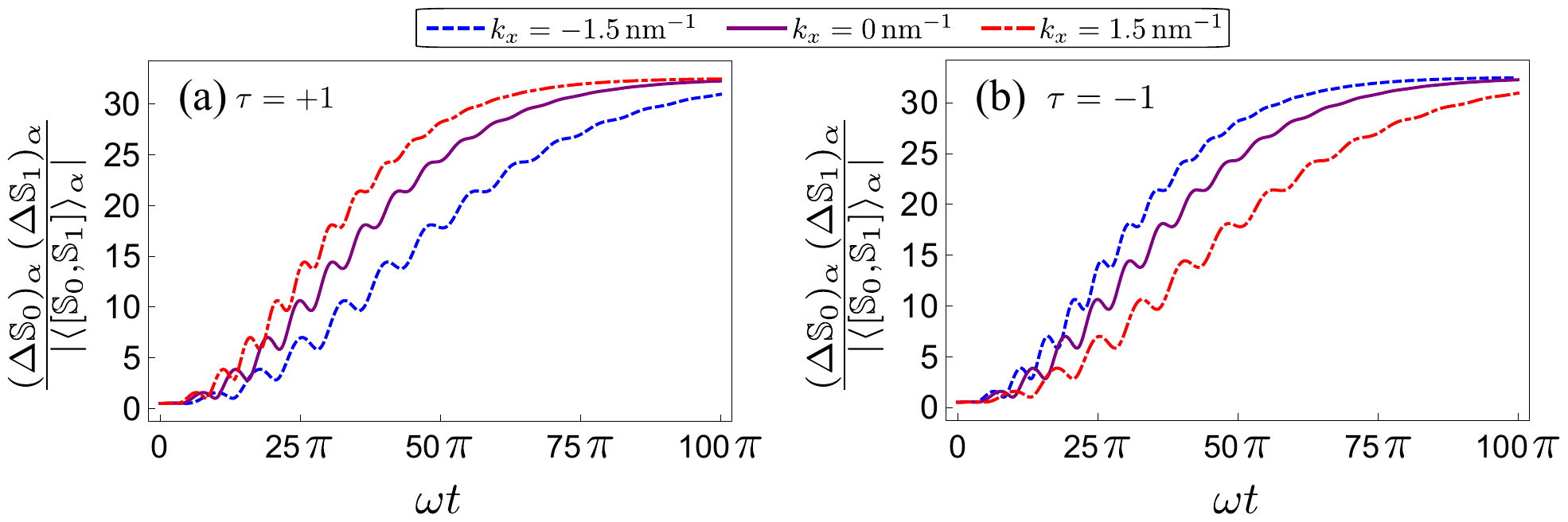}
	\caption{\label{fig:posmomen}Time evolution of the GUP with $\alpha=4$, as function of $\omega\,t$, for $B_{\rm ps}=5$ T and three different values of $k_{x}$: (a) for the valley $\mathbf{K}_{+}$ and (b) for the valley $\mathbf{K}_{-}$.}
\end{figure}

\begin{figure}[ht]
	\centering
	\includegraphics[width=0.9\textwidth]{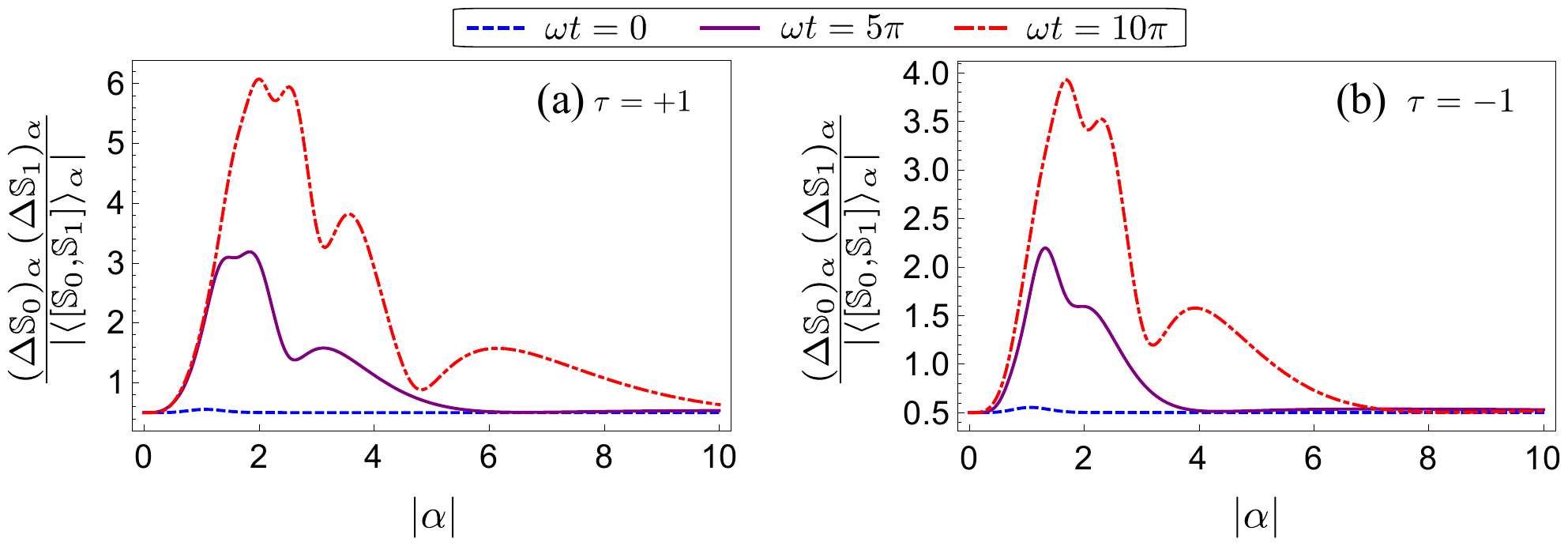}
	\caption{\label{fig:GUP}The GUP with $k_{x}=1.5$ nm$^{-1}$, as function of $\vert\alpha\vert$, for $B_{\rm ps}=5$ T and three different values of $\omega\,t$: (a) for the valley $\mathbf{K}_{+}$ and (b) for the valley $\mathbf{K}_{-}$.}
\end{figure}

Let us consider now the time evolution of the auto-correlation function,
\begin{equation}
    C_{\alpha}(t)=\vert\langle\Psi_{\alpha}\vert\Psi_{\alpha}(t)\rangle\vert=\left(\frac{1}{\exp\left(\vert z\vert^{2}\right)+1}\right)\left\vert1+\sum_{n=0}^{\infty}\frac{\vert z\vert^{2n}}{n!}\exp\left(-\frac{i E_{n,k_{x}}t}{\hbar}\right)\right\vert,
\end{equation}
for studying the time evolution of the PCSs inasmuch as $C(t)$ correlates the same state at two points in time. As Fig.~\ref{fig:autocorrelation} shows, the function $C_{\alpha}(t)$ first oscillates smoothly and afterward faster, with a sinusoidal-like enveloping.

In comparison with the results shown in~\cite{dbe20}, we can see that $k_{x}$ plays a role similar to that of tensile or compression deformations in auto-correlation function, modifying what time the fast oscillations appear, that also depends on the valley index.

Finally, we investigate the time evolution of the generalized uncertainty principle (GUP),
\begin{equation}
    \Delta A\Delta B\geq\frac{1}{2}\vert\langle[A,B]\rangle\vert,
\end{equation}
where $\Delta O^{2}\equiv\langle O^{2}\rangle-(\langle O\rangle)^{2}$ for any operator $O$, by defining the following quadrature
\begin{equation}
    \mathbb{S}_{q}=\frac{1}{\sqrt{2}i^{q}}\left(\Theta^{-}+(-1)^{q}\Theta^{+}\right), \quad q=0,1.
\end{equation}

The mean value of the operator $\mathbb{S}_{q}$ and its square $\mathbb{S}_{q}^{2}$ are given by
\begin{subequations}
\begin{align}
    \langle\mathbb{S}_{q}\rangle_{\alpha}(t)&=\frac{2}{\exp\left(\vert z\vert^{2}\right)+1}\left(\frac{f(z,t,1)+(-1)^{q}f^{\ast}(z,t,1)}{\sqrt{2}i^{q}}\right),\\
    \langle\mathbb{S}_{q}^{2}\rangle_{\alpha}(t)&=\frac{1}{\exp\left(\vert z\vert^{2}\right)+1}\left((2\vert z\vert^{2}+1)\exp\left(\vert z\vert^{2}\right)-\frac{\vert z\vert^{2}}{2}+2(-1)^{q}\Re\left[f(z,t,2)\right]\right),
\end{align}
\end{subequations}
where
\begin{equation}
f(z,t,k)=\frac{z^{k}}{\sqrt{2^{(1-\delta_{2k})}}}\sum_{n=0}^{\infty}\frac{\vert z\vert^{2n}}{n!}\exp\left(-\frac{i(E_{n+k,k_{x}}-E_{n,k_{x}})t}{\hbar}\right), \quad k=1,2.
\end{equation}
Therefore, the GUP for the PCS can be expressed as
\begin{equation}\label{positionmomentum}
    (\Delta\mathbb{S}_{0})_{\alpha}\,(\Delta\mathbb{S}_{1})_{\alpha}\geq\frac{1}{2}\vert\langle[\mathbb{S}_{0},\mathbb{S}_{1}]\rangle_{\alpha}\vert=\frac{1}{\exp\left(\vert z\vert^{2}\right)+1}\left(\exp\left(\vert z\vert^{2}\right)+\frac{\vert z\vert^{2}}{2}\right).
\end{equation}

According to the time evolution of the GUP in Fig.~\ref{fig:posmomen}, as $t$ increases, the uncertainty relation does not keep the minimum value equal to $1/2$. Indeed, the maximum value of the uncertainty is approximately given by $2\vert\alpha\vert^{2}$, whose integral part also corresponds to the Landau level $n$ for which the probability distribution in Appendix~\ref{AppB} reaches its maximum value. Also, similar to the auto-correlation function case, we can see in Fig.~\ref{fig:posmomen} that the momentum $k_{x}$ affects the time evolution of the GUP according to the valley index. Likewise, in Fig.~\ref{fig:GUP} we compare the GUP behaviour in each valley for $\vert\alpha\vert$ growing. For instance, the valley $\mathbf{K}_{-}$ favors that the uncertainty relation keeps minimal for a positive value of the $x$-momentum and $t\neq0$. Besides, by making the identification with the canonical position and momentum operators, namely, $Y\equiv\mathbb{S}_{0}$ and $P_{y}\equiv\mathbb{S}_{1}$, and their mean values $\langle Y\rangle\propto\Re[\alpha]$ and $\langle P_{y}\rangle\propto\Im[\alpha]$, respectively, we can give our coherent states a quite ``classical face'' for describing a quantum mechanical motion in which the position and momentum can be well-defined as might be desired, just by choosing a sufficiently large value of $\vert\alpha\vert$.

\section{Phase-space representation}\label{sec4}
The phase-space formalism is an alternative quantum mechanics formulation whose fundamentals were developed by Groenewold~\cite{Groenewold}, Moyal~\cite{moyal1949}, Weyl~\cite{Weyl1927} and Wigner~\cite{w32}. In this formulation, the standard operator multiplication in quantum mechanics is replaced by the so-called star product~\cite{Groenewold}, which is different in each representation of the phase-space distribution~\cite{zcf05,cz12}. In particular, the so-called Wigner function (WF) is a quasiprobability distribution defined in the phase space formulation of quantum mechanics~\cite{Cahill,berry77,bffls78,igf82}, which appears in many physics branches, e.g., quantum optics~\cite{takahashi86,Knight,Marguerite}, quantum tomography~\cite{Leiner,Knyazev,Jullien}, quantum information processes~\cite{Gu} and electron transport~\cite{Jacoboni,Morandi,Mason,Iafrate,Ferry}. 

Thus, in order to continue the PCSs analysis through a phase-space representation, let us consider the WF $W(\mathbf{q},\mathbf{p})$ defined as~\cite{w32}
\begin{equation}
W(\mathbf{q},\mathbf{p})=\frac{1}{\left(2\pi\hbar\right)^n}\int_{-\infty}^{\infty}\exp\left(\frac{i}{\hbar}\,\mathbf{p}\cdot\mathbf{r}'\right)\left\langle\mathbf{q}-\frac{\mathbf{r}'}{2}\right\vert\rho\left\vert\mathbf{q}+\frac{\mathbf{r}'}{2}\right\rangle d\mathbf{r}',
\label{WM}
\end{equation}
where $\rho$ is the density matrix; $\mathbf{q}=(q_1,q_2,\dots,q_n)$ and $\mathbf{p}=(p_1,p_2,\dots,p_n)$ are vectors defined in a $n$-dimensional space, which represent the classical phase-space position and momentum values, respectively; and $\mathbf{r}'=(r'_1,r'_2,\dots,r'_n)$ is the position vector needed for the integration process. The corresponding normalization condition is given by
\begin{equation}
\int_{-\infty}^{\infty}\int_{-\infty}^{\infty}W(\mathbf{q},\mathbf{p})d\mathbf{q}\,d\mathbf{p}=1.
\end{equation}

\begin{figure}[ht]
	\centering
	\includegraphics[width=0.95\textwidth]{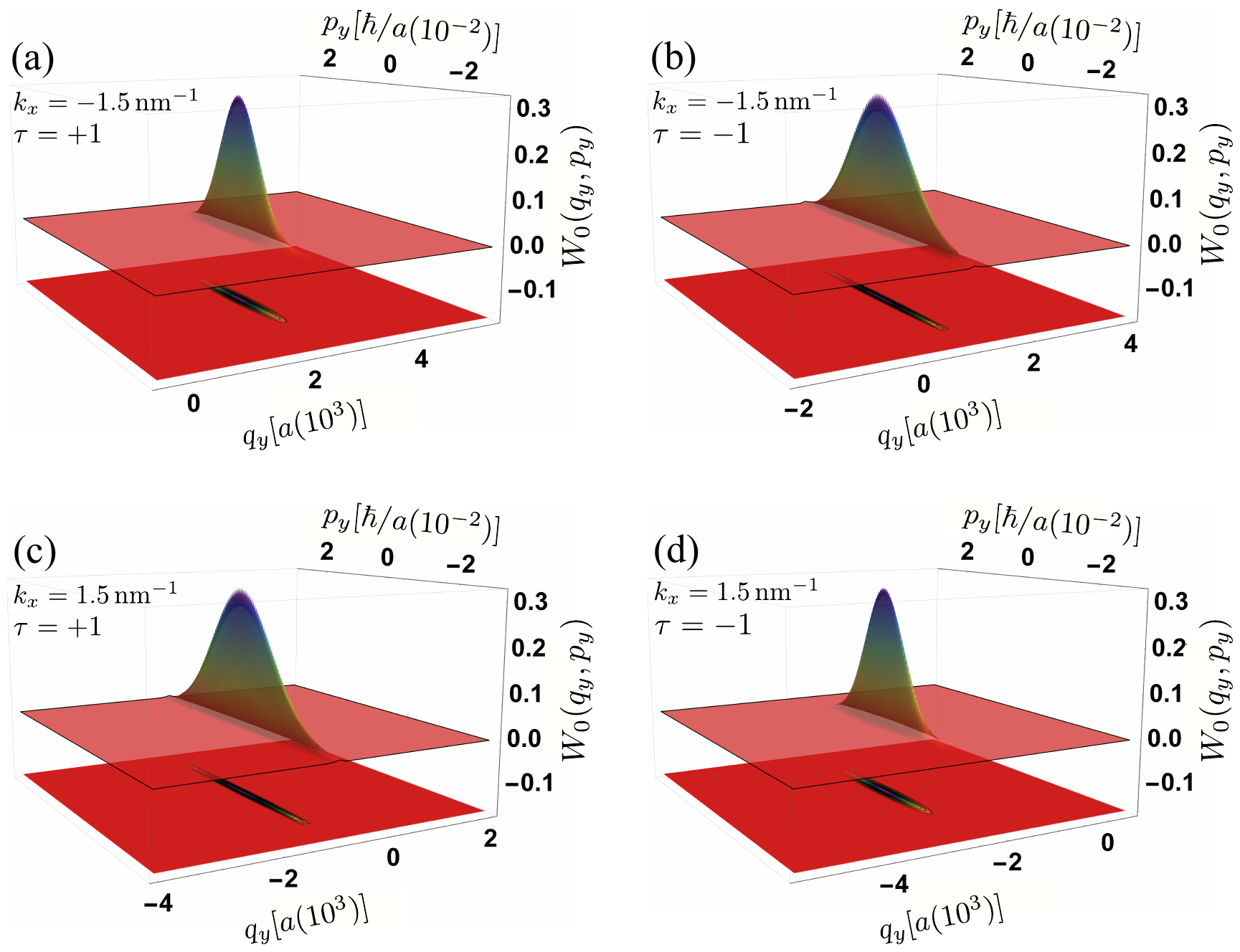}
	\caption{\label{fig:fig6}Trace of the Wigner matrix $W_{n}(\mathbf{q},\mathbf{p})$ in Eq.~(\ref{traceWn}) with $n=0$, for $B_{\rm ps}=5\,\text{T}$ and two different values of $k_{x}$. (a, c) For the valley $\mathbf{K}_{+}$  and (b, d) for the valley $\mathbf{K}_{-}$.}
\end{figure}

\begin{figure}[ht]
	\centering
	\includegraphics[width=0.95\textwidth]{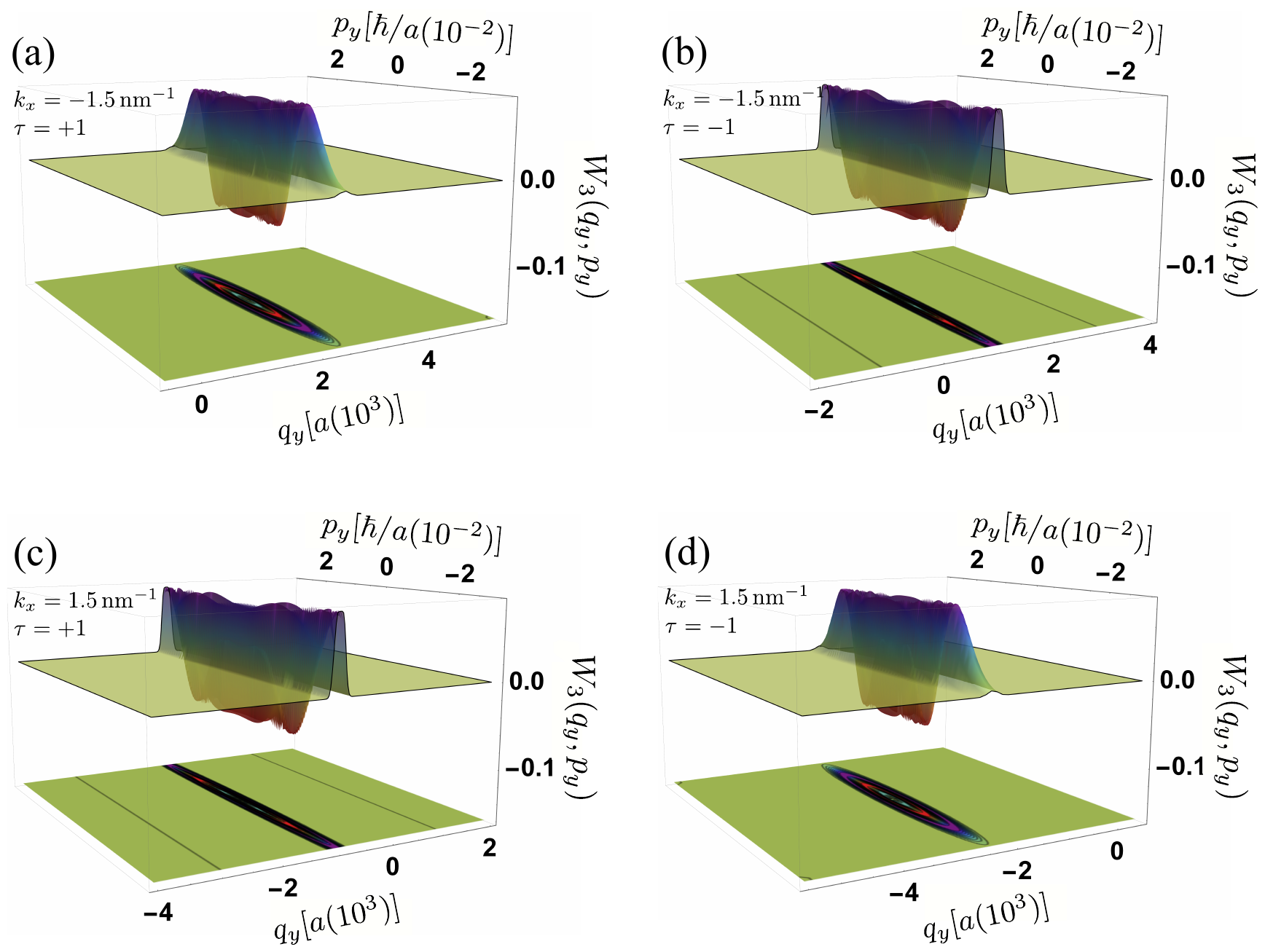}
	\caption{\label{fig:fig7}Trace of the Wigner matrix $W_{n}(\mathbf{q},\mathbf{p})$ in Eq.~(\ref{traceWn}) with $n=3$, for $B_{\rm ps}=5\,\text{T}$ and two different values of $k_{x}$. (a, c) For the valley $\mathbf{K}_{+}$  and (b, d) for the valley $\mathbf{K}_{-}$.}
\end{figure}

Although the WF cannot be considered as a probability distribution since it is a real function that takes negative values, which is an indicator of quantumness and has also been measured experimentally~\cite{Kenfack,w84,r85,sbrf93,bsm97}, its obtaining could establish interesting analogies between the effective Dirac-like approach at low-energy in graphene and quantum optics systems~\cite{diazbetancur20,Jellal,Rusin,Dora,Goldman,Schliemann}.

\subsection{Obtaining of Wigner function}
For the eigenvectors in Eq.~(\ref{eigenvectors}), the corresponding Wigner function, or most properly $2\times2$ Wigner matrix (WM), can be expressed as follows:
\begin{equation}\label{traceWn}
	W_{n}(\mathbf{q},\mathbf{p})=\frac{1}{2^{(1-\delta_{0n})}}W(q_{x},p_{x})\left(\begin{array}{c c}
	W_{n,n}(q_{y},p_{y}) & \tau\,(1-\delta_{0n})W_{n,n-1}(q_{y},p_{y}) \\
	\tau\,(1-\delta_{0n})W_{n-1,n}(q_{y},p_{y}) & (1-\delta_{0n})W_{n-1,n-1}(q_{y},p_{y})
	\end{array}\right),
\end{equation}
where the components $W(q_{x},p_{x})$ and $W_{\mu,\nu}(q_{y},p_{y})$ are given, respectively, by
\begin{subequations}
\begin{align}
		W(q_{x},p_{x})&=\frac{1}{\pi\hbar}\int_{-\infty}^{\infty}\exp\left(2i\left(\frac{p_{x}}{\hbar}-k_{x}\right)z_{1}\right)dz_{1}\equiv\delta\left(p_{x}-k_{x}\hbar\right), \\
		W_{\mu,\nu}(q_{y},p_{y})&=\frac{1}{\pi\hbar}\int_{-\infty}^{\infty}\exp\left(2\frac{i}{\hbar}p_{y}z_2\right)\psi_{\mu}(q_{y}-z_2)\psi_{\nu}^\ast(q_{y}+z_{2}) dz_{2}, \label{18}
\end{align}
\end{subequations}
and $\psi_{\mu}$ and $\psi_{\nu}$ being the wave functions of the quantum harmonic oscillator \eqref{wavefunctions}.

\begin{figure}[ht]
	\centering
	\includegraphics[width=0.95\textwidth]{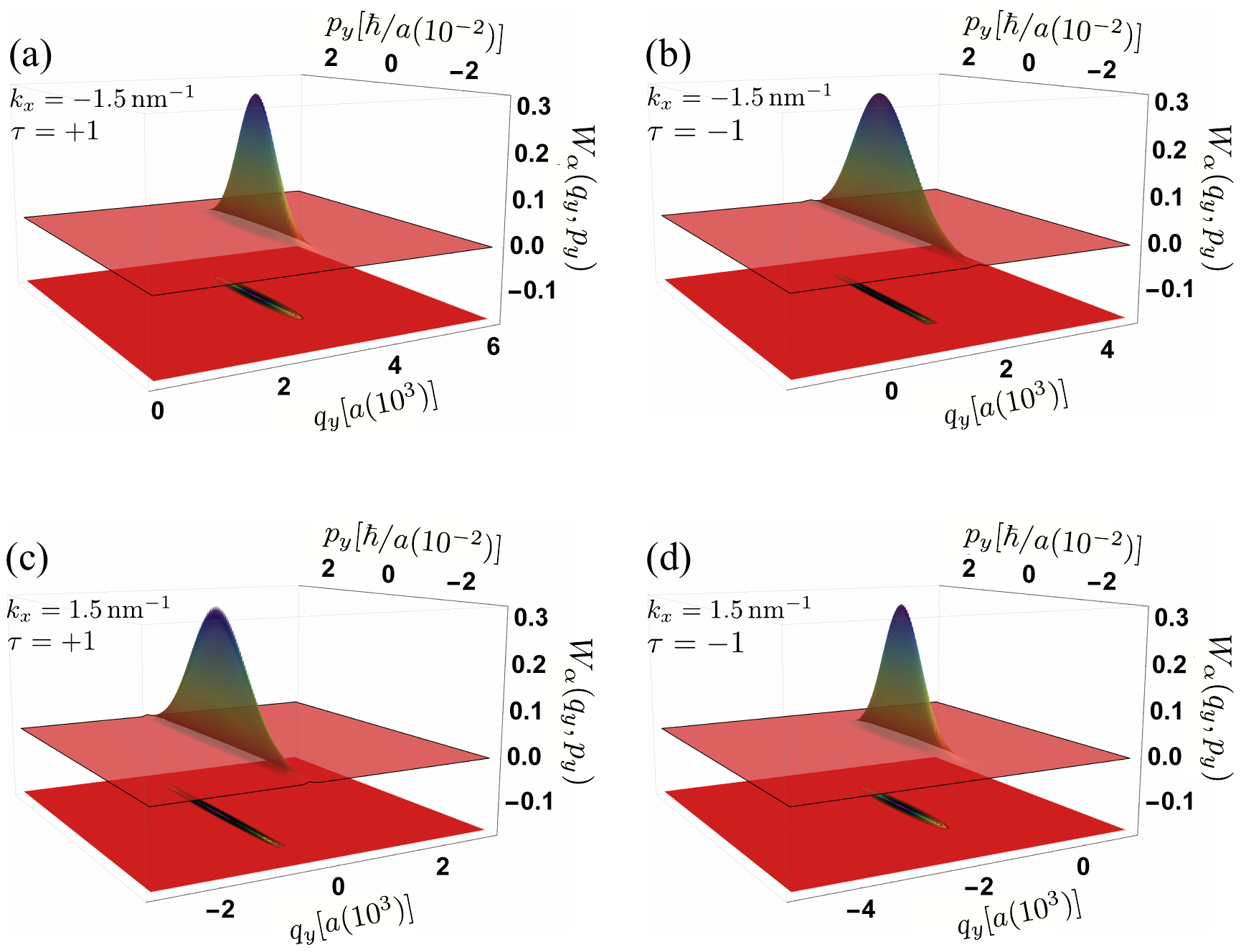}
	\caption{\label{fig:fig8}Trace of the Wigner matrix $W_{\alpha}(\mathbf{q},\mathbf{p})$ in Eq.~(\ref{WFPCS}) with $\alpha=4$, for $B_{\rm ps}=5\,\text{T}$ and two different values of $k_{x}$. (a, c) For the valley $\mathbf{K}_{+}$  and (b, d) for the valley $\mathbf{K}_{-}$.}
\end{figure}

\begin{figure}[ht]
	\centering
	\includegraphics[width=0.95\textwidth]{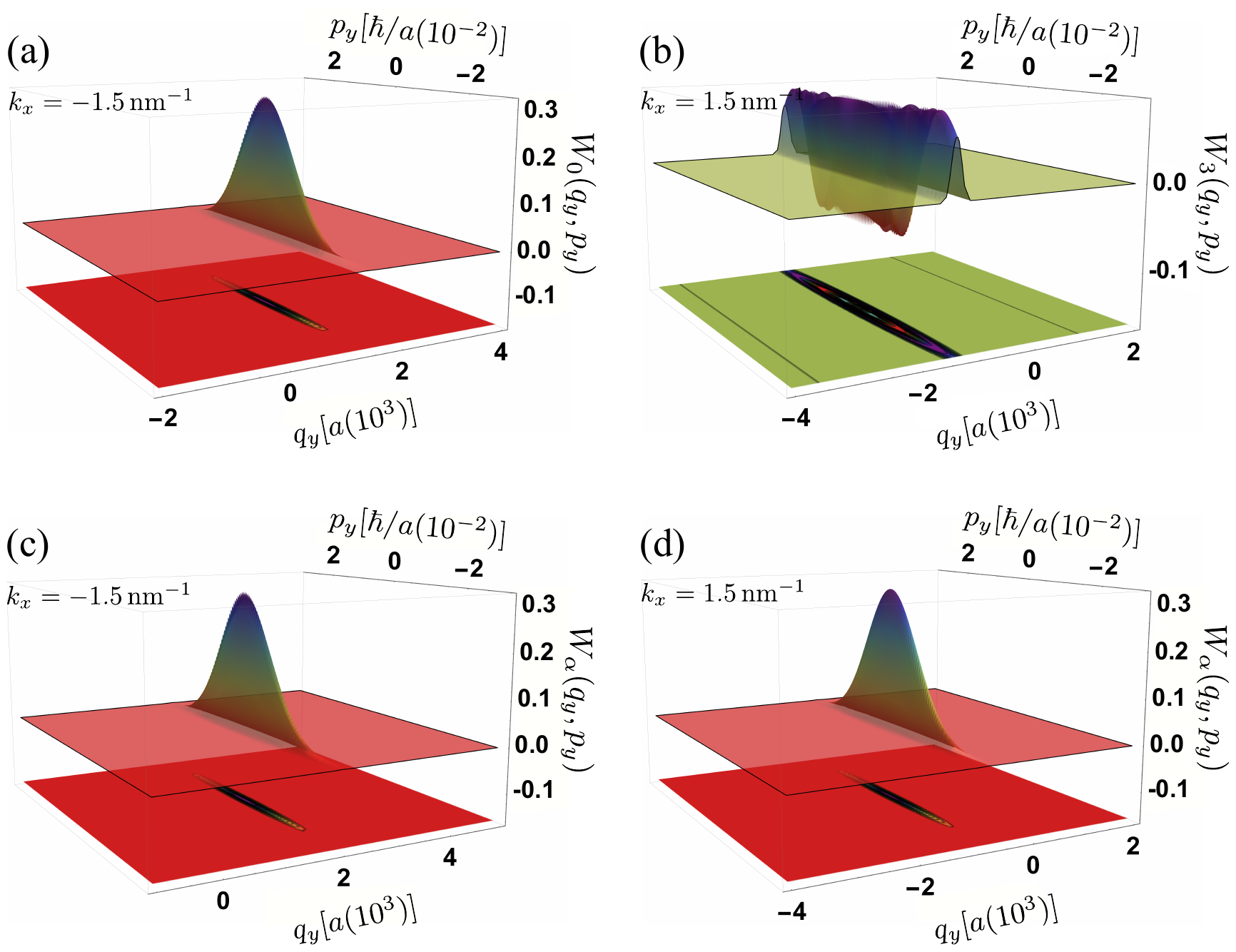}
	\caption{\label{fig:compariosn3}For comparison: Trace of the Wigner matrix $W_{n}(\mathbf{q},\mathbf{p})$ with $n=0$ (a) and $n=3$ (b), and $W_{\alpha}(\mathbf{q},\mathbf{p})$ with $\alpha=4$ (c, d), for different values of $k_{x}$ and an external uniform magnetic of strength $\mathcal{B}_{\rm ext}=5$ T applied to a pristine graphene sample.}
\end{figure}

For computing the function $W_{\mu,\nu}(q_{y},p_{y})$, we define the following quantities
\begin{equation}\label{quantities}
	\xi=\frac{1}{l_{\rm B}}\left(q_{y}+l_{\rm B}^{2}k_{x}\right), \quad y=\frac{z_{2}}{l_{\rm B}}, \quad s=\frac{l_{\rm B}p_{y}}{\hbar}.
\end{equation}
Hence, by substituting Eq.~(\ref{wavefunctions}) in Eq.~(\ref{18}) and using the definitions (\ref{quantities}), we get~\cite{Lev2002}
\begin{equation}\label{38}
	W_{\mu,\nu}(\chi)=\frac{\exp\left(-\frac{1}{2}|\chi|^2\right)}{\pi\hbar}\times\left\{\begin{array}{c c}
		(-1)^\mu\sqrt{\frac{\mu!}{\nu!}}\chi^{\nu-\mu}L_{\mu}^{\nu-\mu}\left(|\chi|^2\right), & \mu\leq\nu, \\
		(-1)^\nu\sqrt{\frac{\nu!}{\mu!}}\chi^{*\mu-\nu}L_{\nu}^{\mu-\nu}\left(|\chi|^2\right), & \mu\geq\nu,
	\end{array}\right.
\end{equation}
where the functions $L_n^m(x)$ are the associated Laguerre polynomials and the quantity $\chi = \sqrt{2}(\xi +is)$ is defined. Thus, the components of the WM turn out to be~\cite{diazbetancur20}
\begin{subequations}\label{23}
	\begin{align}
    W_{n,n}(\chi)&=\frac{(-1)^{n}}{\pi\hbar}\exp\left(-\frac{1}{2}|\chi|^2\right)L_{n}\left(|\chi|^2\right),\\
	W_{n-1,n}(\chi)&=W_{n,n-1}^\ast(\chi)=\frac{(-1)^{n-1}}{\pi\hbar\sqrt{n}}\chi \exp\left(-\frac{1}{2}|\chi|^2\right)L_{n-1}^{1}\left(|\chi|^2\right), \\
	W_{n-1,n-1}(\chi)&=\frac{(-1)^{n-1}}{\pi\hbar}\exp\left(-\frac{1}{2}|\chi|^2\right)L_{n-1}\left(|\chi|^2\right).
	\end{align}
\end{subequations}

Figures~\ref{fig:fig6} and~\ref{fig:fig7} show the trace of the WM associated to the fundamental state $\Psi_{0}(\mathbf{r})$ and excited state $\Psi_{3}(\mathbf{r})$, respectively. As occurs for the HO, the WF associated to the ground state is positive in whole phase space, while for the excited states, it takes negative values.

On the other hand, the WM for the PCSs is given by
\begin{equation}\label{WFPCS}
W_{\alpha}(\mathbf{q},\mathbf{p})=\frac{\delta\left(p_{x}-k_{x}\hbar\right)}{2\left(\exp\left(\vert z\vert^2\right)+1\right)}\left(\begin{array}{c c}
W_{11}(q_{y},p_{y}) & \tau\,W_{12}(q_{y},p_{y}) \\
\tau\,W_{21}(q_{y},p_{y}) & W_{22}(q_{y},p_{y})
\end{array}\right),
\end{equation}
where
\begin{subequations}\label{componentsWF}
\begin{align}
W_{11}(\chi)&=\frac{1}{\pi\hbar}\exp\left(-\frac{1}{2}\vert\chi\vert^{2}\right)\left\{1+\exp\left(\vert\chi\vert^{2}-\vert\chi- z\vert^{2}\right)+2\Re\left[\exp\left( z\chi^{\ast}\right)\right]\right\}, \\
W_{12}(\chi)&=W_{21}^{\ast}(\chi)=\frac{1}{\pi\hbar}\exp\left(-\frac{1}{2}|\chi|^2\right)\sum_{m=1}^{\infty}\frac{( z^*)^{m}}{m!}\sqrt{m}\left(\chi^{m-1}+\exp\left( z\chi^*\right)\left(\chi- z\right)^{m-1}\right), \\
W_{22}(\chi)& = -\frac{\textrm{e}^{-\frac{1}{2}|\chi|^2}}{\pi\hbar}\sum_{n=1}^{\infty}\left[\frac{(-| z|^2)^n}{n!}L_{n-1}\left(|\chi|^2\right) +2\sum_{m=n+1}^{\infty}\frac{(-1)^{n}}{m!}\Re\{z^nz^{\ast m}\chi^{m-n}\}\sqrt{\frac{m}{n}}L_{n-1}^{m-n}\left(|\chi|^2\right)\right]. \label{44c}
\end{align}
\end{subequations}

Figure~\ref{fig:fig8} shows the trace of the WM for the PCSs. Similarly to that of harmonic-oscillator coherent states, the function $W_{\alpha}(\mathbf{q},\mathbf{p})$ is also positive in whole phase space. However, the WM for the PCSs also shown some differences. First, its location along the $q_{y}$-axis depends on the value of the momentum $k_{x}$. This is related to that the center of the cyclotron in the Landau-like gauge is given by~\cite{dbe20}
\begin{equation}\label{cyclotron}
    X_{0}=x+\frac{l_{\rm B}^{2}p_{y}}{\hbar}, \quad Y_{0}=y-\frac{l_{\rm B}^{2}\left(p_{x}+e\mathcal{B}y\right)}{\hbar}=-\frac{l_{\rm B}^{2}p_{x}}{\hbar},
\end{equation}
i.e., the wave functions in  Eq.~(\ref{eigenvectors}) are eigenstates of $Y_{0}$ with eigenvalue $q_{y}=-l_{\rm B}^{2}k_{x}=-\hbar k_{x}/(eB_{\rm ps}(1+\tau2k_{x}a))$, which is precisely where the WM is located along the $q_{y}$-axis ($\xi=0$ in Eq.~(\ref{quantities})). Second, the WF width changes according to the valley index and the value of the momentum $k_{x}$, which is not observed even in the uniaxially-deformed case~\cite{diazbetancur20}. Therefore, the $x$-momentum dependency is not trivial, as occurs when an external magnetic field $\mathbf{B}_{\rm ext}$ is applied (see Fig.~\ref{fig:compariosn3}), and its effects are observed in the time evolution of coherent states and the behavior of their corresponding WM.

\section{Conclusions}\label{conclusions}
The semi-classical dynamics of low-energy quasiparticles in 2D Dirac materials has been studied through the construction of coherent states in previous works. These quantum states have been built as eigenstates of a matrix annihilation operator corresponding to the low-energy Hamiltonian for graphene (either pristine or uniformly deformed) under an external homogeneous magnetic field.

In this work, we have considered a nonuniform  honeycomb lattice where the nearest hopping parameters are position dependent (such as occur in strained graphene). As a consequence, the low-energy quasiparticles of this system are affected by a pseudomagnetic field and a position-dependent Fermi velocity and, therefore, they exhibit no longer flat but dispersive pseudo-Landau levels. In order to describe the semi-classical dynamics of Dirac fermions with such an unusual Landau level spectrum, we have based the coherent state construction on a different approach with respect to the previous works: after defining matrix ladder operators $\Theta^{\pm}$, we considered a non-unitary displacement operator $D(\alpha)$, whose action on the fundamental eigenvector $\Psi_{0}(\mathbf{r})$ allowed obtaining of CSs, $\Psi_{\alpha}(\mathbf{r})$, as a linear combination of the eigenvectors $\Psi_{n}(\mathbf{r})$ for $\lambda=+$. From these quantum states, we were able to analyze and describe the effects of having a pseudomagnetic field and a position-dependent Fermi velocity on the dynamics of Dirac fermions. 
According to our results, the position-dependent Fermi velocity in Eq.~(\ref{tensorvelocity}), which induces both a $k_{x}$-momentum and a valley-index dependency in the Landau levels, affects the time evolution of the PCSs. More precisely, the period of motion $T$ changes depending on the value of the momentum $k_{x}$: for charged particles that belong to the Dirac point $\mathbf{K}_{+}$ ($\mathbf{K}_{-}$), $T$ increases (decreases) when $k_{x}$ takes negative values, while $T$ decreases (increases) when $k_{x}$ is positive, as shown in Figs.~\ref{fig:density} and \ref{fig:autocorrelation}. Likewise, the time evolution of the GUP is modulated by the valley index, the $x$-momentum and the parameter $\alpha$, as Figs.~\ref{fig:posmomen} and \ref{fig:GUP} show, in order to keep minimal uncertainty. More precisely, for $\vert\alpha\vert\gg1$, a $\Psi_{\alpha}$ state describes very well the motion of a macroscopic oscillator, for which the position, the momentum and the energy can be considered to be classical quantities. Also, the width of the WF and its location along the $q_{y}$-axis are also affected by the value of $k_{x}$, in contrast with the case of uniformly strained graphene, where the momentum in the orthogonal direction only affects the location, not the shape. These results constitute remarkable differences with respect to the previous works~\cite{dcr19,diazbetancur20,diaz2019coherent,dbe20} for 2D Dirac materials with an anisotropic but homogeneous Fermi velocity under a external magnetic field. In general, our findings for dispersive pseudo-Landau-levels expand the CS formalism and offer a novel scenario for electronic transport studios.

\section*{Acknowledgments}
This work was supported by Consejo Nacional de Ciencia y Tecnología (Mexico), project FORDECYT-PRONACES/61533/2020 and Secretaría de Investigación y Posgrado (Instituto Politécnico Nacional) Grant 20210317.

\appendix
\numberwithin{equation}{section}
\section{Orthogonality and completeness relation}\label{AppA}
The PCSs satisfy the following relation
\begin{equation}
    \vert\langle\Psi_{z'}\vert\Psi_{z}\rangle\vert=\left\vert\frac{\exp(z'^{\ast}z)+1}{\sqrt{(\exp(\vert z\vert^{2})+1)(\exp(\vert z'\vert^{2})+1)}}\right\vert\neq\delta(z'-z).
\end{equation}
This implies that these states are not orthogonal for $z\neq z'$, so we can say that the set of states $\Psi_{\alpha}$ is overcomplete.

On the other hand, it is worth to remark that the PCS do not satisfy a completeness relation in the usual sense, since the superposition considers positive energy states only~\cite{roy21}. In order to clarify this, let us consider the following expression:
\begin{equation}\label{complete}
    \int_{\mathbb{C}}\vert\Psi_{z}\rangle\langle\Psi_{z}\vert\frac{d\mu(z)}{\pi}+\frac12\sum_{n=1}^{\infty}\vert\Psi_{n}\rangle\langle\Psi_{n}\vert,
\end{equation}
where $d\mu(z)$ is the measure defined in the complex plane as
\begin{equation}
    d\mu(z)=\frac{\vert z\vert\left(\exp\left(\vert z\vert^{2}\right)+1\right)}{2\exp\left(\vert z\vert^{2}\right)}d\vert z\vert\,d\theta.
\end{equation}
Now, by defining the variable $r=\vert z\vert$, Eq.~(\ref{complete}) can be rewritten as:
\begin{align}
 \nonumber   &\int_{0}^{\infty}\int_{0}^{2\pi}\frac{e^{-r^{2}}r}{4\pi}\left[\vert\Psi_{0}\rangle\langle\Psi_{0}\vert+\sum_{m=0}^{\infty}\frac{r^{m}e^{-im\theta}}{\sqrt{m!}}\vert\Psi_{0}\rangle\langle\Psi_{m}\vert+\sum_{n=0}^{\infty}\frac{r^{n}e^{in\theta}}{\sqrt{n!}}\vert\Psi_{n}\rangle\langle\Psi_{0}\vert\right.\\
    &\quad\left.+\sum_{n,m=0}^{\infty}\frac{r^{n+m}e^{i(n-m)\theta}}{\sqrt{n!\,m!}}\vert\Psi_{n}\rangle\langle\Psi_{m}\vert\right] d\theta dr+\frac12\sum_{n=1}^{\infty}\vert\Psi_{n}\rangle\langle\Psi_{n}\vert,
\end{align}
which, after applying the results 
\begin{equation}
    \int_{0}^{2\pi}\exp\left(i(n-m)\theta\right)d\theta=2\pi\delta_{mn}, \quad \int_{0}^{\infty}2r^{2n+1}\exp\left(-r^{2}\right)dr=\Gamma(n+1)=n!,
\end{equation}
yields to
\begin{equation}
    \sum_{n=0}^{\infty}\vert\Psi_{n}\rangle\langle\Psi_{n}\vert\equiv\mathbb{I},
\end{equation}
where $\mathbb{I}$ denotes the identity operator in the Hilbert space $\mathcal{H}$ of Landau levels in the conduction band ($\lambda=+$).

\section{Occupation number distribution}\label{AppB}
On the other hand, the probability of a PCS of being in an eigenstate $\Psi_{n}$ is given by
\begin{equation}\label{Poisson}
    P_{\alpha}(n)=\vert\langle\Psi_{n}\vert\Psi_{\alpha}\rangle\vert^{2}=\left(\frac{2}{\exp\left(2\mu\right)+1}\right)\times\begin{cases}
    1, & n=0, \\
    \frac{(2\mu)^{n}}{2n!}, & n>0,
    \end{cases}
\end{equation}
where $\mu=\vert\alpha\vert^{2}$.

\begin{figure}[ht]
	\centering
	\includegraphics[width=8cm]{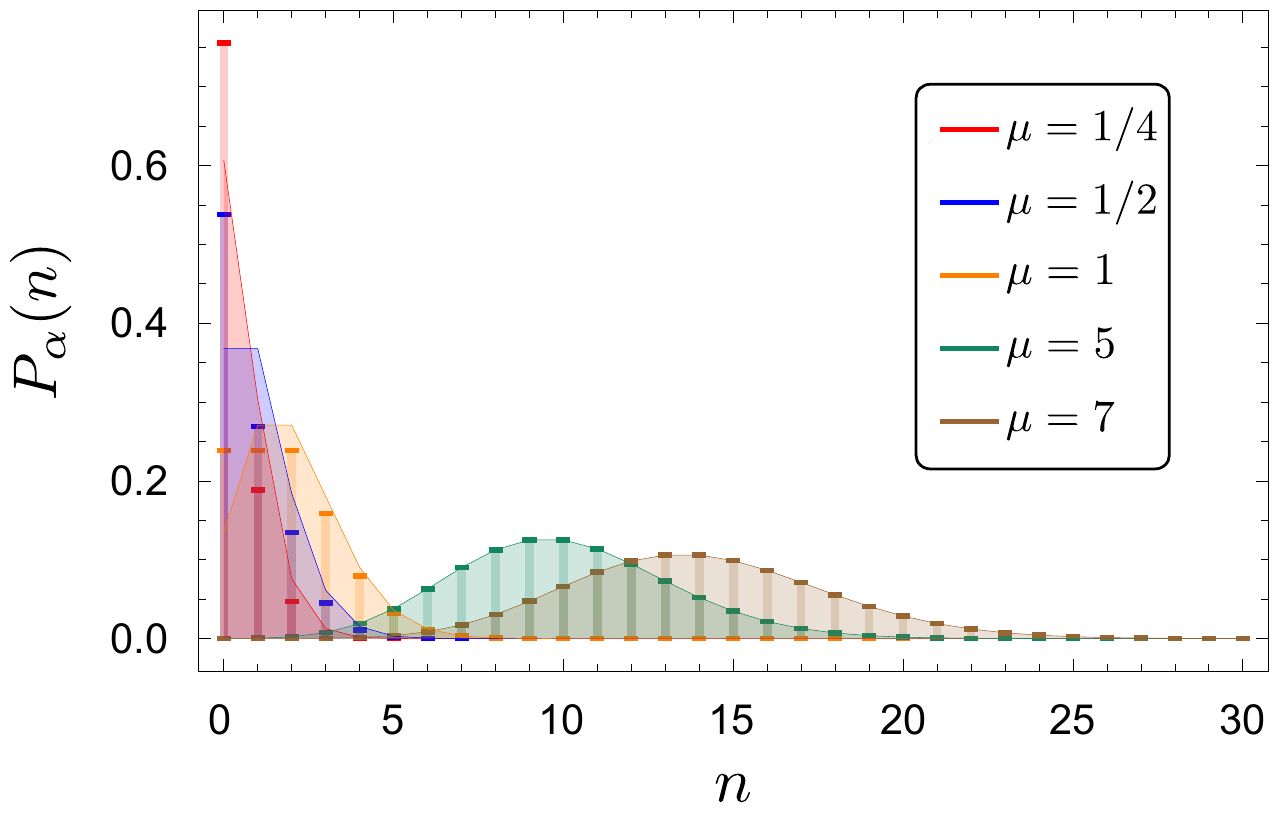}
	\caption{\label{fig:Poisson}Occupation number distribution $P_{\alpha}(n)$ in Eq.~(\ref{Poisson}) for the coherent states $\Psi_{\alpha}$ for different values of $\mu=\vert\alpha\vert^{2}$. $P_{\alpha}(n)$ adjusts to Poisson distribution (solid curves) as $\mu$ grows.}
\end{figure}

This occupation number distribution is compared with that of the CSs of the harmonic oscillator with eigenvalue $z_{\rm CS}\in\mathbb{C}$ in Fig.~\ref{fig:Poisson}. For the harmonic oscillator case, $P_{z}(n)$ is a Poisson distribution, namely,  $P_{z}(n)=\exp(-\tau)\tau^{n}/n!$ with mean $\tau=\vert z_{\rm CS}\vert^{2}$. In our case, as $\mu$ increases, we have that $P_{\alpha}(n)\sim P_{z}(n)$ with $\tau=2\mu$.

\bibliographystyle{ieeetr}
\bibliography{biblio}

\end{document}